\documentclass[jap,showpacs,twocolumn,groupedaddress]{revtex4}

\usepackage{graphicx}

\def\ensuretext{\textrm}
\newcommand{\fig}[4][h]{\begin{figure}[#1]\begin{center}\includegraphics[scale=#2]{#3.pdf}\vspace{-0.25 cm}\caption{#4}\label{fig:#3}\end{center}\end{figure}}

\newcommand{\ket}[1]{\ensuremath{\left|{#1}\right\rangle}}

\newcommand{\Rb}[1]{\ensuremath{^{#1}}\ensuretext{Rb}}
\newcommand{\hf}[2]{\ket{#1, \; #2}}
\newcommand{\units}[1]{\ensuretext{\thinspace #1}}

\newcommand{\etal}[0]{\emph{et al.}}

\begin{document}

\title{Measurement of inelastic losses in a sample of ultracold \Rb{85}}

\author{P. A. Altin, N. P. Robins, R. Poldy, J. E. Debs, D. D\"oring, C. Figl and J. D. Close}
\affiliation{Australian Centre for Quantum Atom Optics, Australian National University, ACT 0200, Australia}
\email{paul.altin@anu.edu.au}
\homepage{http://atomlaser.anu.edu.au/}

\date{\today}

\begin{abstract}

We report on the observation and characterisation of an inelastic loss feature in collisions between ultracold \Rb{85} \hf{F=2}{m_F=-2} atoms at a magnetic field of $220\units{G}$. Our apparatus creates ultracold \Rb{85} clouds by sympathetic cooling with a \Rb{87} reservoir, and can produce pure \Rb{87} condensates of $10^6$ atoms by a combination of evaporative cooling in a quadrupole-Ioffe magnetic trap and further evaporation in a weak, large-volume optical dipole trap. By combining \Rb{85} and \Rb{87} atoms collected in a dual-species magneto-optical trap and selectively evaporating the heavier isotope, we demonstrate strong sympathetic cooling of the \Rb{85} cloud, increasing its phase space density by three orders of magnitude with no detectable loss in number. We have used ultracold samples created in this way to observe the variation of inelastic loss in ultracold \Rb{85} as a function of magnetic field near the $155\units{G}$ Feshbach resonance. We have also measured a previously unobserved loss feature at $219.9(1)\units{G}$ with a width of $0.28(6)\units{G}$, which we associate with a narrow Feshbach resonance predicted by theory.

\end{abstract}

\pacs{34.50.-s, 37.10.De, 37.10.Gh, 67.85.Hj, 67.85.Jk}

\maketitle

The ability to tune the interparticle interactions in an atomic Bose-Einstein condensate (BEC) has opened up a wide range of new experiments in the field of ultracold atoms. Many experiments have used a Feshbach resonance, a magnetically-tunable molecular bound state, to modify the $s$-wave scattering properties of ultra-cold atoms. The most notable of these include the Bose-Einstein condensation of cesium \cite{weber02}, the formation of a molecular BEC from a Fermi gas \cite{regal04}, and the demonstration of atom interferometry with a weakly-interacting condensate \cite{fattori07}. Recent proposals for generating squeezing and massive particle entanglement in an atom laser would also benefit from the ability to tune the $s$-wave elastic scattering length of the atom beam \cite{haine06,johnsson07}.

The first tunable-interaction condensate was of \Rb{85}, created by Cornish \etal\ at JILA in 2000 \cite{cornish00}. This atom is well-suited to studying a wide range of interaction strengths, with a broad, low field Feshbach resonance that allows the $s$-wave scattering length to be tuned over several orders of magnitude, including both positive and negative values \cite{cornish00}. However, inelastic loss processes\textemdash and in particular the variation of these losses with magnetic field\textemdash make condensation of \Rb{85} difficult to achieve. Indeed, it was only through detailed knowledge of these processes and their behaviour near the Feshbach resonance \cite{roberts00} that a successful path to condensation was devised by the JILA group.

In this paper, we report on a measurement of a previously unobserved increase in inelastic loss associated with a narrow Feshbach resonance in \Rb{85} near $220\units{G}$ \cite{burke99}. We have also made a second measurement of the losses near the $155\units{G}$ Feshbach resonance, observing losses in agreement with the theoretical predictions of Refs. \cite{burke98,roberts00}. To obtain an ultracold sample for study, we employ sympathetic cooling with a reservoir of \Rb{87} in a magnetic trap. This technique relies on the large interspecies collision cross-section between \Rb{87} and \Rb{85} \cite{burke98}. It was first applied to \Rb{85} by Bloch \etal\ \cite{bloch01}, and has more recently been used to create stable \Rb{85} condensates by Papp \etal\ \cite{papp06}. By operating at a minimum in the inelastic loss rate we are able to further evaporate \Rb{85} in an optical dipole trap to phase-space densities consistent with BEC.

In our experiment, $10^{10}$ \Rb{87} atoms and $10^8$ \Rb{85} atoms are collected simultaneously in an ultra-high vacuum three-dimensional magneto-optical trap (MOT), fed by a cold atom beam from a two-dimensional MOT. The flux from the 2D MOT is sufficient to fill the 3D MOT within $10\units{s}$. After a $20\units{ms}$ polarization-gradient cooling stage, the MOT repumping lasers are switched off for $1\units{ms}$ to allow both species to be optically pumped into their lower ground states in preparation for sympathetic cooling. The atoms are captured in a quadrupole magnetic field and magnetically transported over a distance of $40\units{mm}$ to a second quadrupole trap, which is then converted into a harmonic Ioffe-Pritchard potential using the quadrupole-Ioffe coil configuration \cite{esslinger98}. The final magnetic potential has trapping frequencies of $\omega_z = 2\pi\times16\units{Hz}$ axially and $\omega_\rho = 2\pi\times156\units{Hz}$ radially for \Rb{87} atoms in the \hf{F=1}{m_F=-1} hyperfine state, and a bias field of $3.7\units{G}$. Typically, $10^9$ \Rb{87} atoms and $10^7$ \Rb{85} atoms are present at $\sim200\units{$\mu$K}$ in our QUIC trap before evaporation. We reduce the temperature of the samples by applying a logarithmic radiofrequency (rf) sweep from $50\units{MHz}$ to $3\units{MHz}$ over $15\units{s}$. The number and temperature of either species at the end of an experimental run is determined from absorption imaging after a period of free expansion.

A combination of poor elastic and inelastic scattering cross-sections makes \Rb{85} badly suited to direct evaporative cooling \cite{burke98,roberts00}. An alternative approach is to cool the sample sympathetically through thermal contact with an evaporated reservoir. In order to effect sympathetic cooling of the \Rb{85} \hf{F=2}{m_F=-2} state, which has a broad Feshbach resonance at $155\units{G}$, the refrigerant species must be preferentially removed from the magnetic trap during evaporation. As demonstrated in 2007 by Papp \etal, this can be achieved using the standard rf forced evaporation technique, provided that the \Rb{87} atoms occupy the \hf{F=1}{m_F=-1} state. Because the Land\'e factor $g_F$ has a larger magnitude for the \Rb{87} ground state ($g_F = \pm1/2$) than for \Rb{85} ($g_F = \pm1/3$), the \Rb{87} evaporation surface at a given radiofrequency is closer to the centre of the trap than for \Rb{85}. In addition, the \Rb{85} cloud is more tightly confined (i.e. smaller) in the magnetic trap due to its larger $g_F\;m_F$ factor. These effects combine to make rf evaporation highly selective towards the coolant \Rb{87} \footnote{The difference in the gravitational sag for each species in the magnetic trap is much smaller than the size of the cloud at temperatures above $\sim1\units{$\mu$K}$, and thus does not contribute to isotope selectivity during evaporation.}.

We have observed strong sympathetic cooling with almost perfect isotope selectivity during evaporation with both species in the magnetic trap. Figure \ref{fig:sympathetic} shows evaporation trajectories for each species, both alone and in the presence of the other. Direct evaporation of \Rb{85} fails due to the low elastic collision rate\textemdash a consequence of the low initial density of the sample and of the small elastic collision cross-section of \Rb{85} at temperatures above $100\units{$\mu$K}$ \cite{burke98}. With both species in the trap, however, no loss of \Rb{85} is detected during cooling from $200\units{$\mu$K}$ to $20\units{$\mu$K}$, corresponding to an increase in phase space density of over three orders of magnitude. The sympathetic cooling trajectory of \Rb{85} begins to roll off at around $20\units{$\mu$K}$, as the number of each species present becomes comparable. This is to be expected, since beyond this point the density of \Rb{85} at the evaporation surface will not necessarily be lower than that of \Rb{87}, and so the rf begins to remove both species from the trap. 

In the work of Bloch \etal\ \cite{bloch01}, sympathetic cooling of \Rb{85} \hf{F=3}{m_F=3} atoms was demonstrated using \Rb{87} in the \hf{F=2}{m_F=2} state. In that system the two species experienced identical confinement, and preferential removal of \Rb{87} was due solely to the larger energy splitting between neighboring Zeeman sublevels (larger $g_F$). This reduced selectivity manifested in the reported \Rb{85} loss of a factor of $\sim3$. In contrast to the \hf{F=2}{m_F=-2} state used in the present work, the \hf{F=3}{m_F=3} state in \Rb{85} has no known Feshbach resonance. The data presented in Figure \ref{fig:sympathetic} are in excellent qualitative agreement with the more recent work of Papp \etal\ (cf. Ref. \cite{papp07}, Fig. 4.11), including the accelerated loss of \Rb{87} that occurs as the number ratio approaches unity.

%%%%
\fig[t]{0.55}{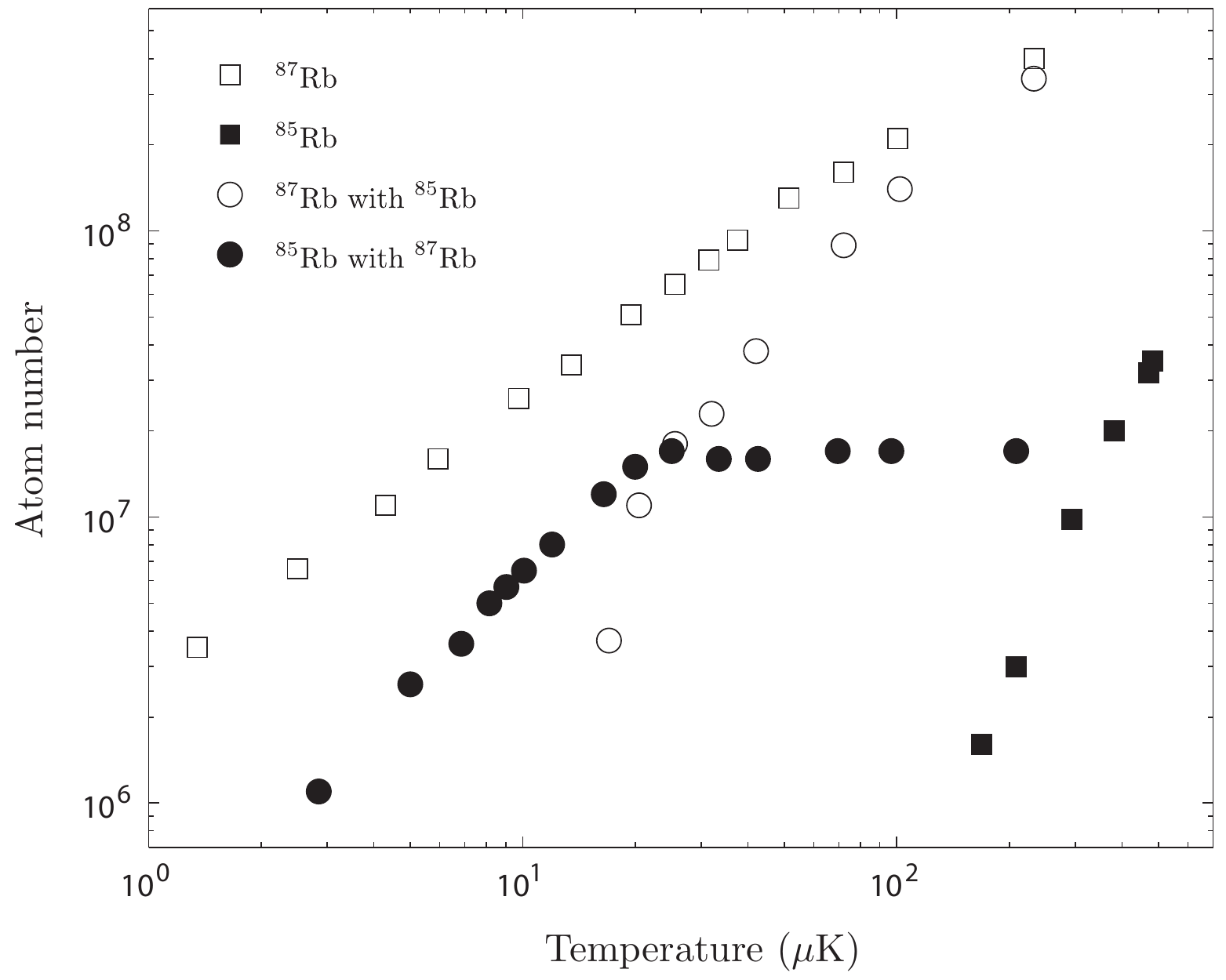}{Atom number as a function of temperature during rf-induced evaporative cooling in the magnetic trap. The evaporation trajectories of \Rb{85} (circles) and \Rb{87} (squares) are shown during both single-species (open) and sympathetic cooling (filled).}
%%%%

Under optimal experimental conditions, we can create samples of $8\times10^6$ \Rb{85} atoms at $10\units{$\mu$K}$ in the QUIC trap, giving a phase space density of $6\times10^{-4}$. With a larger \Rb{87} reservoir, it may be possible to continue the sympathetic cooling further. However, the inelastic scattering properties of \Rb{85} are known to prevent the containment of high density samples at low magnetic fields\textemdash it is expected that sympathetic cooling would no longer be effective below $\sim5\units{$\mu$K}$, depending on the density of the sample \cite{papp07}. Even without inelastic losses, a \Rb{85} BEC at low field would be unstable against collapse with more than $\sim100$ atoms \cite{ruprecht95} due to the negative background scattering length, $a \approx -440a_0$. Both of these issues may be overcome by exploiting the Feshbach resonance between ultracold \Rb{85} \hf{F=2}{m_F=2} atoms which occurs at $155\units{G}$.

The radial confinement of a QUIC trap decreases with bias field as $\omega_\rho \propto B_0^{-1/2}$, so that at high fields the radial confinement of our trap would be insufficient to support the atoms against gravity. Thus the magnetic trap is not suitable for investigating the behavior of an ultracold \Rb{85} cloud as a function of magnetic field around the $155\units{G}$ Feshbach resonance. For this we use an optical dipole trap formed by focusing $3\units{W}$ of light from a $1090\units{nm}$ fibre laser to a $100\units{$\mu$m}$ waist. The weak axes and radial oscillation frequencies of the two traps are matched, and we transfer the atoms from the magnetic trap to the dipole trap by suddenly ($<100\units{$\mu$s}$) superimposing the dipole laser onto the magnetically-trapped cloud. A bias field of around $160\units{G}$ produced by a pair of Helmholtz coils along the beam axis is then ramped up over $500\units{ms}$. This destroys the radial confinement of the magnetic trap, however a small magnetic field curvature along the beam (corresponding to an oscillation frequency of $\sim 10\units{Hz}$) is left on to strengthen the axial confinement, which would otherwise be only $0.2\units{Hz}$ along the beam axis. Finally, we evaporate by lowering the power in the dipole beam, thus allowing energetic atoms to escape radially from the trap. In the absence of \Rb{85} atoms, we are able to produce pure \Rb{87} condensates containing more than $10^6$ atoms in the \hf{F=1}{m_F=-1} state.

%%%%
\fig[t]{0.5}{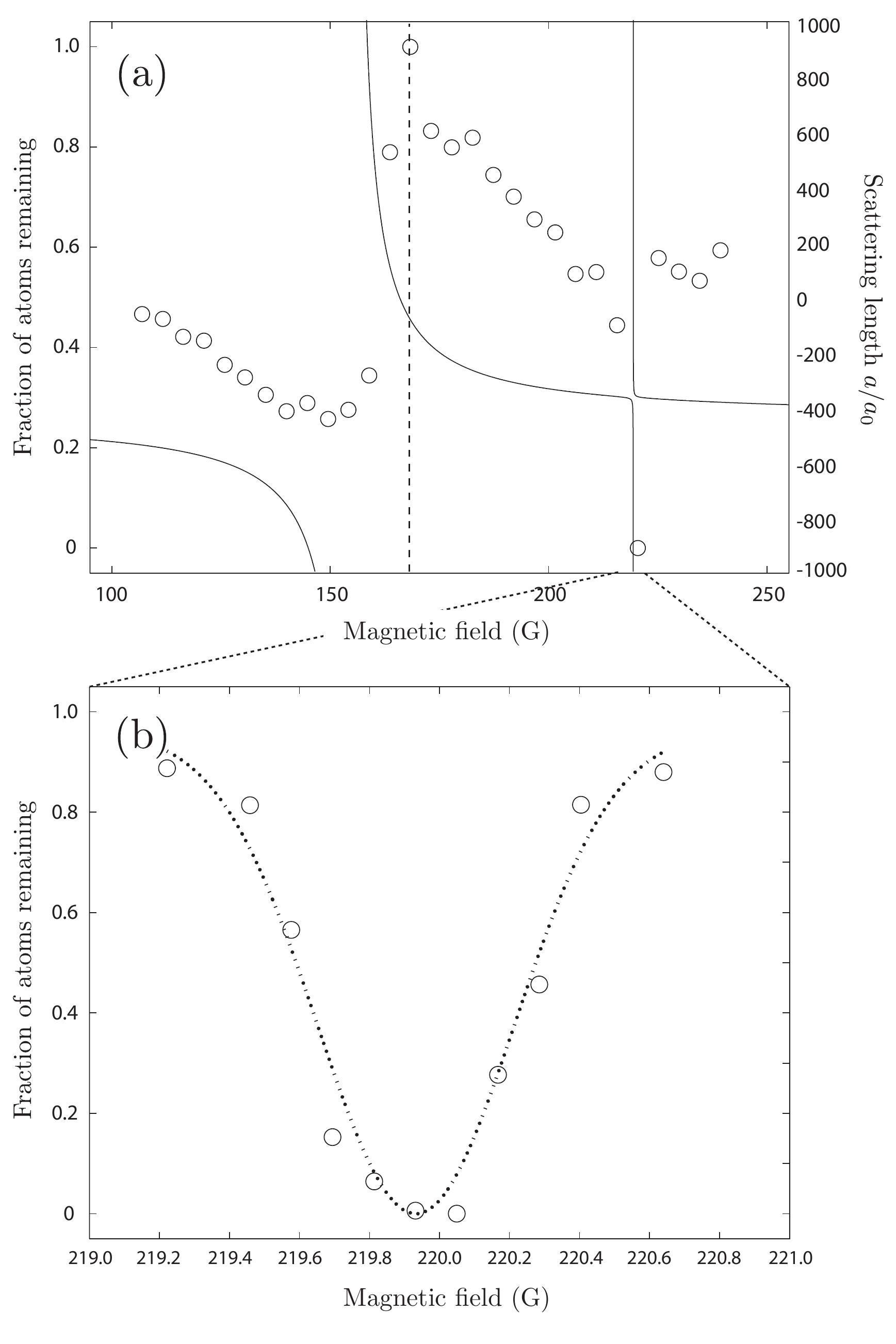}{(a) Measured inelastic loss of \Rb{85} from the optical dipole trap as a function of magnetic field, with the elastic scattering length from Ref. \cite{kohler06} overlaid (solid line). The minimum inelastic loss occurs at $168\units{G}$ (dashed line). \protect\linebreak (b) Zoomed-in section showing a sharp increase in inelastic loss associated with a Feshbach resonance at $220\units{G}$. The dashed line is a Gaussian fit to the experimental data.}
%%%%

To investigate the behaviour of inelastic processes in \Rb{85}, we load a sympathetically-cooled sample from the QUIC trap into the optical trap and ramp up the bias field over $500\units{ms}$. The \Rb{85} cloud initially present in the optical dipole trap has a density of $\sim3\times10^{12}\units{cm$^{-3}$}$ and a temperature of $2.5\units{$\mu$K}$. After a $10\units{s}$ hold at the desired field, the atoms are released from the trap and imaged. The bias field is calibrated by subjecting the cloud to a burst of radiofrequency; since the axial confinement is magnetic, atoms which transition to a high-field seeking state are expelled from the trap. Measuring loss as a function of applied rf frequency therefore allows us to determine the magnetic field at the cloud centre using the Breit-Rabi equation \cite{breit31}. Figure \ref{fig:feshbach}(a) shows the number of atoms remaining in the trap after $10\units{s}$ as a function of the applied bias field. Loss from the dipole trap is enhanced on the low field side of the Feshbach resonance and peaks as the elastic scattering length diverges at $155\units{G}$. The inelastic loss is minimised at around $168\units{G}$, where the scattering length vanishes, consistent with results reported by the JILA group \cite{cornish00,papp06}. Overall, the curve shows good qualitative agreement with the theoretical predictions of inelastic loss rates by Burke \etal\ \cite{burke98} (cf. Fig. 2) and Roberts \etal\ \cite{roberts00} (cf. Fig. 2b).

We also observe a sharp increase in inelastic loss near $220\units{G}$. Theoretical calculations \cite{burke99} predict the existence of a narrow Feshbach resonance at this field, with the most recent placing the resonance at $219.4\units{G}$ with a width of $\sim10\units{mG}$ \cite{julienne09}. Our observed loss feature (Fig. \ref{fig:feshbach}(b)) occurs at $(219.9\pm0.1)\units{G}$. We expect no significant systematic error due to light shifts from the dipole laser; since the detuning is much larger than the separation between the $m_F$ states, each magnetic sublevel is shifted by the same amount, thus both the calibration of the magnetic field by rf addressing and the position of the resonance itself should be unaffected by the optical trap (to first order). The loss at this field is significantly greater than at the broader Feshbach resonance, even though the two-body loss coefficient is predicted to be two orders of magnitude lower here than at $155\units{G}$ \cite{roberts00}, implying a particularly high three-body loss rate.

The width of the curve in Fig. \ref{fig:feshbach}(b), $\sigma_{\text{meas}} = (0.28\pm0.06)\units{G}$, is a convolution of the width of the inelastic loss feature $\sigma_{\text{inel}}$ and the effective width of the cloud $\sigma_{\text{cloud}}$ due to the range of magnetic fields spanned by the atoms, power broadening of the rf transition, and the magnetic field noise. These effects are reduced by using the coldest possible cloud, minimal rf power and low current noise power supplies to drive the Feshbach bias coils. We measure the frequency width of the cloud to be $(100\pm10)\units{kHz}$, corresponding to a magnetic field width of $\sigma_{\text{cloud}} = (55\pm5)\units{mG}$. Thus we deduce the width of the inelastic loss feature to be $\sigma_{\text{inel}} = (0.27\pm0.07)\units{G}$. This should not be expected to match the width of the resonance in the \emph{elastic} scattering cross-section.

As shown in Figure \ref{fig:feshbach}(a), inelastic losses between \Rb{85} \hf{F=2}{m_F=-2} atoms are minimised at around $168\units{G}$, near where the $s$-wave scattering length vanishes. Operating at this magnetic field, we have further cooled both species in the optical trap by evaporation. Since the \Rb{85} $s$-wave scattering length at this field is much lower than the interspecies scattering length, this cooling is again almost purely sympathetic, although at the end of the ramp the scattering length is tuned to a larger positive value to make the formation of a condensate possible. By tuning the relative numbers of \Rb{87} and \Rb{85} loaded into the dipole trap, we can optimise the cooling performance to achieve the highest phase space density in \Rb{85}. In this manner, we have produced ultracold \Rb{85} clouds containing $10^4$ atoms at $\sim10\units{nK}$ coexisting with condensates of \Rb{87} \footnote{We have recently achieved Bose-Einstein condensation of \Rb{85} in this machine, using the method outlined above. This will be described in detail in an upcoming paper, along with a comparison of our machine to the \Rb{85} BEC machine at JILA.}.

In conclusion, we have used a sample of ultracold \Rb{85} \hf{F=2}{m_F=-2} atoms to measure loss due to inelastic collisions as a function of magnetic field around the $155\units{G}$ Feshbach resonance, and to characterise a previously unobserved inelastic loss feature at $220\units{G}$. We have demonstrated efficient sympathetic cooling of \Rb{85} atoms in a QUIC magnetic trap using a \Rb{87} reservoir\textemdash an important step in creating a \Rb{85} Bose-Einstein condensate. The phase space density of more than $10^7$ \Rb{85} atoms was increased by three orders of magnitude without loss through thermal contact with an evaporated \Rb{87} cloud. We have further cooled \Rb{85} in an optical dipole trap to phase space densities consistent with BEC.

The authors gratefully acknowledge G. Dennis and C. Savage for helpful discussions and B. Buchler for experimental assistance. This work is supported by the Australian Research Council Centre of Excellence for Quantum-Atom Optics.

\bibliography{mk2}

\begin{thebibliography}{18}
\expandafter\ifx\csname natexlab\endcsname\relax\def\natexlab#1{#1}\fi
\expandafter\ifx\csname bibnamefont\endcsname\relax
  \def\bibnamefont#1{#1}\fi
\expandafter\ifx\csname bibfnamefont\endcsname\relax
  \def\bibfnamefont#1{#1}\fi
\expandafter\ifx\csname citenamefont\endcsname\relax
  \def\citenamefont#1{#1}\fi
\expandafter\ifx\csname url\endcsname\relax
  \def\url#1{\texttt{#1}}\fi
\expandafter\ifx\csname urlprefix\endcsname\relax\def\urlprefix{URL }\fi
\providecommand{\bibinfo}[2]{#2}
\providecommand{\eprint}[2][]{\url{#2}}

\bibitem[{\citenamefont{Weber et~al.}(2002)\citenamefont{Weber, Herbig, Mark,
  N{\"a}gerl, and Grimm}}]{weber02}
\bibinfo{author}{\bibfnamefont{T.}~\bibnamefont{Weber}},
  \bibinfo{author}{\bibfnamefont{J.}~\bibnamefont{Herbig}},
  \bibinfo{author}{\bibfnamefont{M.}~\bibnamefont{Mark}},
  \bibinfo{author}{\bibfnamefont{H.-C.} \bibnamefont{N{\"a}gerl}},
  \bibnamefont{and} \bibinfo{author}{\bibfnamefont{R.}~\bibnamefont{Grimm}},
  \bibinfo{journal}{Science} \textbf{\bibinfo{volume}{299}},
  \bibinfo{pages}{232} (\bibinfo{year}{2002}).

\bibitem[{\citenamefont{Regal et~al.}(2004)\citenamefont{Regal, Greiner, and
  Jin}}]{regal04}
\bibinfo{author}{\bibfnamefont{C.}~\bibnamefont{Regal}},
  \bibinfo{author}{\bibfnamefont{M.}~\bibnamefont{Greiner}}, \bibnamefont{and}
  \bibinfo{author}{\bibfnamefont{D.}~\bibnamefont{Jin}},
  \bibinfo{journal}{Phys. Rev. Lett.} \textbf{\bibinfo{volume}{92}},
  \bibinfo{pages}{040403} (\bibinfo{year}{2004}).

\bibitem[{\citenamefont{Fattori et~al.}(2007)\citenamefont{Fattori, {D'Errico},
  Roati, Zaccanti, {Jona-Lasinio}, Modugno, Inguscio, and Modugno}}]{fattori07}
\bibinfo{author}{\bibfnamefont{M.}~\bibnamefont{Fattori}},
  \bibinfo{author}{\bibfnamefont{C.}~\bibnamefont{{D'Errico}}},
  \bibinfo{author}{\bibfnamefont{G.}~\bibnamefont{Roati}},
  \bibinfo{author}{\bibfnamefont{M.}~\bibnamefont{Zaccanti}},
  \bibinfo{author}{\bibfnamefont{M.}~\bibnamefont{{Jona-Lasinio}}},
  \bibinfo{author}{\bibfnamefont{M.}~\bibnamefont{Modugno}},
  \bibinfo{author}{\bibfnamefont{M.}~\bibnamefont{Inguscio}}, \bibnamefont{and}
  \bibinfo{author}{\bibfnamefont{G.}~\bibnamefont{Modugno}},
  \bibinfo{journal}{Phys. Rev. Lett.} \textbf{\bibinfo{volume}{100}},
  \bibinfo{pages}{080405} (\bibinfo{year}{2007}).

\bibitem[{\citenamefont{Johnsson and Haine}(2007)}]{johnsson07}
\bibinfo{author}{\bibfnamefont{M.}~\bibnamefont{Johnsson}} \bibnamefont{and}
  \bibinfo{author}{\bibfnamefont{S.}~\bibnamefont{Haine}},
  \bibinfo{journal}{Phys. Rev. Lett.} \textbf{\bibinfo{volume}{99}},
  \bibinfo{pages}{010401} (\bibinfo{year}{2007}).

\bibitem[{\citenamefont{Haine et~al.}(2006)\citenamefont{Haine, Olsen, and
  Hope}}]{haine06}
\bibinfo{author}{\bibfnamefont{S.}~\bibnamefont{Haine}},
  \bibinfo{author}{\bibfnamefont{M.}~\bibnamefont{Olsen}}, \bibnamefont{and}
  \bibinfo{author}{\bibfnamefont{J.}~\bibnamefont{Hope}},
  \bibinfo{journal}{Phys. Rev. Lett.} \textbf{\bibinfo{volume}{96}},
  \bibinfo{pages}{133601} (\bibinfo{year}{2006}).

\bibitem[{\citenamefont{Cornish et~al.}(2000)\citenamefont{Cornish, Claussen,
  Roberts, Cornell, and Wieman}}]{cornish00}
\bibinfo{author}{\bibfnamefont{S.}~\bibnamefont{Cornish}},
  \bibinfo{author}{\bibfnamefont{N.}~\bibnamefont{Claussen}},
  \bibinfo{author}{\bibfnamefont{J.}~\bibnamefont{Roberts}},
  \bibinfo{author}{\bibfnamefont{E.}~\bibnamefont{Cornell}}, \bibnamefont{and}
  \bibinfo{author}{\bibfnamefont{C.}~\bibnamefont{Wieman}},
  \bibinfo{journal}{Phys. Rev. Lett.} \textbf{\bibinfo{volume}{85}},
  \bibinfo{pages}{1795} (\bibinfo{year}{2000}).

\bibitem[{\citenamefont{Roberts et~al.}(2000)\citenamefont{Roberts, Claussen,
  Cornish, and Wieman}}]{roberts00}
\bibinfo{author}{\bibfnamefont{J.}~\bibnamefont{Roberts}},
  \bibinfo{author}{\bibfnamefont{N.}~\bibnamefont{Claussen}},
  \bibinfo{author}{\bibfnamefont{S.}~\bibnamefont{Cornish}}, \bibnamefont{and}
  \bibinfo{author}{\bibfnamefont{C.}~\bibnamefont{Wieman}},
  \bibinfo{journal}{Phys. Rev. Lett.} \textbf{\bibinfo{volume}{85}},
  \bibinfo{pages}{728} (\bibinfo{year}{2000}).

\bibitem[{\citenamefont{{Burke, Jr.}}(1999)}]{burke99}
\bibinfo{author}{\bibfnamefont{J.}~\bibnamefont{{Burke, Jr.}}},
  \bibinfo{journal}{PhD thesis, University of Colorado}
  (\bibinfo{year}{1999}),
  \urlprefix\url{http://jilawww.colorado.edu/pubs/thesis/burke/}.

\bibitem[{\citenamefont{{Burke, Jr.} et~al.}(1998)\citenamefont{{Burke, Jr.},
  Bohn, Esry, and Greene}}]{burke98}
\bibinfo{author}{\bibfnamefont{J.}~\bibnamefont{{Burke, Jr.}}},
  \bibinfo{author}{\bibfnamefont{J.}~\bibnamefont{Bohn}},
  \bibinfo{author}{\bibfnamefont{B.}~\bibnamefont{Esry}}, \bibnamefont{and}
  \bibinfo{author}{\bibfnamefont{C.}~\bibnamefont{Greene}},
  \bibinfo{journal}{Phys. Rev. Lett.} \textbf{\bibinfo{volume}{80}},
  \bibinfo{pages}{2097} (\bibinfo{year}{1998}).

\bibitem[{\citenamefont{Bloch et~al.}(2001)\citenamefont{Bloch, Greiner,
  Mandel, H{\"a}nsch, and Esslinger}}]{bloch01}
\bibinfo{author}{\bibfnamefont{I.}~\bibnamefont{Bloch}},
  \bibinfo{author}{\bibfnamefont{M.}~\bibnamefont{Greiner}},
  \bibinfo{author}{\bibfnamefont{O.}~\bibnamefont{Mandel}},
  \bibinfo{author}{\bibfnamefont{T.}~\bibnamefont{H{\"a}nsch}},
  \bibnamefont{and}
  \bibinfo{author}{\bibfnamefont{T.}~\bibnamefont{Esslinger}},
  \bibinfo{journal}{Phys. Rev. Lett.} \textbf{\bibinfo{volume}{87}},
  \bibinfo{pages}{030401} (\bibinfo{year}{2001}).

\bibitem[{\citenamefont{Papp and Wieman}(2006)}]{papp06}
\bibinfo{author}{\bibfnamefont{S.}~\bibnamefont{Papp}} \bibnamefont{and}
  \bibinfo{author}{\bibfnamefont{C.}~\bibnamefont{Wieman}},
  \bibinfo{journal}{Phys. Rev. Lett.} \textbf{\bibinfo{volume}{97}},
  \bibinfo{pages}{180404} (\bibinfo{year}{2006}).

\bibitem[{\citenamefont{Esslinger et~al.}(1998)\citenamefont{Esslinger, Bloch,
  and H{\"a}nsch}}]{esslinger98}
\bibinfo{author}{\bibfnamefont{T.}~\bibnamefont{Esslinger}},
  \bibinfo{author}{\bibfnamefont{I.}~\bibnamefont{Bloch}}, \bibnamefont{and}
  \bibinfo{author}{\bibfnamefont{T.}~\bibnamefont{H{\"a}nsch}},
  \bibinfo{journal}{Phys. Rev. A} \textbf{\bibinfo{volume}{58}},
  \bibinfo{pages}{R2664} (\bibinfo{year}{1998}).

\bibitem[{\citenamefont{Papp}(2007)}]{papp07}
\bibinfo{author}{\bibfnamefont{S.}~\bibnamefont{Papp}}, \bibinfo{journal}{PhD
  thesis, University of Colorado}  (\bibinfo{year}{2007}),
  \urlprefix\url{http://jilawww.colorado.edu/pubs/thesis/papp/}.

\bibitem[{\citenamefont{Ruprecht et~al.}(1995)\citenamefont{Ruprecht, Holland,
  Burnett, and Edwards}}]{ruprecht95}
\bibinfo{author}{\bibfnamefont{P.}~\bibnamefont{Ruprecht}},
  \bibinfo{author}{\bibfnamefont{M.}~\bibnamefont{Holland}},
  \bibinfo{author}{\bibfnamefont{K.}~\bibnamefont{Burnett}}, \bibnamefont{and}
  \bibinfo{author}{\bibfnamefont{M.}~\bibnamefont{Edwards}},
  \bibinfo{journal}{Phys. Rev. A} \textbf{\bibinfo{volume}{51}},
  \bibinfo{pages}{4704} (\bibinfo{year}{1995}).

\bibitem[{\citenamefont{K\"{o}hler et~al.}(2006)\citenamefont{K\"{o}hler,
  G\'{o}ral, and Julienne}}]{kohler06}
\bibinfo{author}{\bibfnamefont{T.}~\bibnamefont{K\"{o}hler}},
  \bibinfo{author}{\bibfnamefont{K.}~\bibnamefont{G\'{o}ral}},
  \bibnamefont{and} \bibinfo{author}{\bibfnamefont{P.}~\bibnamefont{Julienne}},
  \bibinfo{journal}{Rev. Mod. Phys.} \textbf{\bibinfo{volume}{78}},
  \bibinfo{pages}{1311} (\bibinfo{year}{2006}).

\bibitem[{\citenamefont{Breit and Rabi}(1931)}]{breit31}
\bibinfo{author}{\bibfnamefont{G.}~\bibnamefont{Breit}} \bibnamefont{and}
  \bibinfo{author}{\bibfnamefont{I.}~\bibnamefont{Rabi}},
  \bibinfo{journal}{Phys. Rev.} \textbf{\bibinfo{volume}{38}},
  \bibinfo{pages}{2082} (\bibinfo{year}{1931}).

\bibitem[{\citenamefont{Julienne}(2009)}]{julienne09}
\bibinfo{author}{\bibfnamefont{P.}~\bibnamefont{Julienne}},
  \bibinfo{howpublished}{private communication} (\bibinfo{year}{2009}).

\bibitem[{\citenamefont{Jochim et~al.}(2003)\citenamefont{Jochim, Bartenstein,
  Altmeyer, Hendl, Chin, {Hecker Denschlag}, and Grimm}}]{jochim03}
\bibinfo{author}{\bibfnamefont{S.}~\bibnamefont{Jochim}},
  \bibinfo{author}{\bibfnamefont{M.}~\bibnamefont{Bartenstein}},
  \bibinfo{author}{\bibfnamefont{A.}~\bibnamefont{Altmeyer}},
  \bibinfo{author}{\bibfnamefont{G.}~\bibnamefont{Hendl}},
  \bibinfo{author}{\bibfnamefont{C.}~\bibnamefont{Chin}},
  \bibinfo{author}{\bibfnamefont{J.}~\bibnamefont{{Hecker Denschlag}}},
  \bibnamefont{and} \bibinfo{author}{\bibfnamefont{R.}~\bibnamefont{Grimm}},
  \bibinfo{journal}{Phys. Rev. Lett.} \textbf{\bibinfo{volume}{91}},
  \bibinfo{pages}{240402} (\bibinfo{year}{2003}).

\end{thebibliography}

\end{document}